%% file: saim_may2015.tex
\documentclass[sunil1,ChapterTOCs]{sunil} %See documentation for other class options
\usepackage{fixltx2e,fix-cm}
\usepackage{amssymb}
\usepackage{amsmath}
\usepackage{graphicx}
\usepackage{subfigure}
\usepackage{makeidx}
\usepackage{multicol}

\newcommand{\out}[1]{}
\begin{document}

\author{
\hspace*{4cm}\parbox{10cm}{\large 
To appear in: \\Luigia Petre and Emil Sekerinski (editors) \\From Action Systems to Distributed Systems: 
\\The Refinement Approach. 
\\CRC Press, Taylor \&\ Francis Group
}\vspace{3cm}\\{\LARGE \textit{Gheorghe Stefanescu}}
}
\title{Self-assembling interactive modules: A research programme} 
\maketitle

\input{chapters/chapterGS/stefmac-kaisa2014.sty}
\input{chapters/chapterGS/struc-2dim-lang_final-may2015}

\bibliographystyle{plain}
\bibliography{bibtex_GS}

\end{document}

%% file: chapters/chapterGS/struc-2dim-lang_final-may2015.tex
% Book K Sere, Chapter Stefanescu

\chapter{Self-assembling interactive modules: A research programme}

In this paper we propose a research programme for getting structural characterisations for 2-dimensional languages generated by self-assembling tiles. This is part of a larger programme on getting a formal foundation of parallel, interactive, distributed systems.

\section{Tiling - a brief introduction}

Tiling is an old and popular subject \cite{book-tiling87}. While our focus is on 2-dimensional tiles/tiling and 2-dimensional regular expressions, the formalism itself can be easily extended to cope with 3, 4, or more dimensions. 

An interesting tiling problem was proposed by Wang in 1961.  \textit{Wang tiles} are: (1) finite sets of unit squares, with a color on each side; (2) which can not be rotated or reflected; and (3) such that there is an infinite number of copies from each tile. A \textit{tiling} is a side-by-side arrangement of tiles, such that neighbouring tiles have the same color on the common borders. The main problem of interest here is the following: \textit{``Given a set of tiles, is there a tiling of the whole 2-dimensional plane?''} Wang has conjectured that only regular periodic tilings are possible, but this conjecture was refuted later and aperiodic tilings of the plane were found; the smallest known set for which an aperiodic tiling does exist consists in 13 tiles \cite{kc,jk}. See also \cite{enp} for some recent results, seen from a mathematical perspective.

Another interesting model using tiles is the Winfree’s abstract Tile Assembly Model; see \cite{pat} for a recent survey. This is a theoretic model aiming to capture the basic features of self-assembling systems from physics, chemistry, or biology. The main problem of interest here is: \textit{``Find a set of tiles such that any tiling yields the same specified final configuration.''} Reformulated in rewriting language terms, the problem of interest is to find confluent and terminating sets of tiling rules. The interest in this problem comes from practical reasons: to use self-assembling for producing complex substances with no errors and in a fast way.

The approach presented in this paper comes form a different perspective: how to use tiling to describe the syntax and the semantics of open, interactive, distributed programs and computing systems. The problem of interest here is: \textit{``Find all finite tiling configurations with a given color on each west/north/east/south border''}. This is an abstract formulation of the basic fact that we are interested in running scenarios of distributed systems, corresponding to tiling configurations, which start from initial states (on north), initial interaction classes (on west) and, in a finite number of steps, reach final states (on south) and final interaction classes (on east). To conclude, while formally our tiles are somehow similar with Wang or Winfree tiles, the problem of interest is different. 

Tiling has also been used for the study of 2-dimensional languages\foo{This approach considers an abstract notion of dimension. Nevertheless, our work on tiling originates from a study of interactive systems (i.e., the register-voice interactive systems model \cite{ste04,ste06a}), leading to a model with 2 dimensions and a particular interpretation of them: the vertical dimension represents \textit{time}, while the horizontal dimension is used for \textit{space}. More on this distinction between abstract tiling and developing scenarios out of interactive modules is included in the last section.}, a topics closer to our approach here. The field of 2-dimensional languages started in 1960s, mostly related to ``picture languages''. The field has got a renewed interest in 1990s, when a robust class of \textit{regular} 2-dimensional languages has been identified; good surveys from that period are \cite{gi-re97,lmn98}. Comparisons between our approach and some known results in the area of 2-dimensional languages are included in the text below, as well as in our previous papers on a new type of 2-dimensional regular expressions to be introduced in the next section, see, e.g., \cite{bps13,ba-st14b,ba-st14a}.

We end this brief introduction with a comment on the use of ``self-assembling'' term here. According to some conventions, a (chemical) self-assembling system is an assembling system with the following distinctive features related to the \textit{order}, the \textit{interactions}, and the used \textit{building blocks}: (1) usually, the resulting configurations have higher order; (2) they use ``weak interactions'' for coordination; and (3) larger or heterogeneous building blocks may be used. A property as (1) is present in the statement of the proposed research programme presented in Section 1.3. Property (3) is a basic ingredient in our new type of 2-dimensional regular expressions, based on scenario composition. Property (2) is more related to physical systems - a slightly similar one may be considered in our setting, making a difference between computing (on a machine) and coordinating activities. In short, there are strong enough reasons to use the ``self-assembling'' term for constructing scenarios in the current distributed computing formalism. Both, the term and the self-assembling way of thinking may be useful when considering distributed software services, as well.

\section{Two-dimensional languages: local vs. global glueing constraints}

\subsection{Words and languages, in two dimensions}

A \textit{2-dimensional word} is a finite area of unit cells, in the lattice $\mathbb{Z}\times\mathbb{Z}$, labelled with letters from a finite alphabet. A \textit{2-dimensional letter} is a 2-dimensional word consisting of a unique cell. A \textit{2-dimensional language} is a set of 2-dimensional words.

\begin{figure}
\begin{tabular}{c}
\includegraphics[scale=0.2]{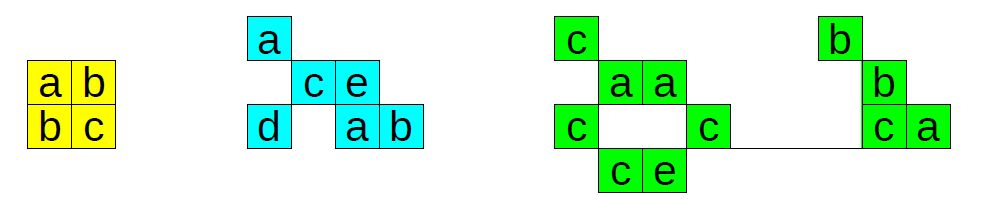}\\
\hspace*{.8cm}(a)\hspace{1.2cm}(b)\hspace{2.5cm}(c)\hspace*{2.cm}
\end{tabular}
\caption{2-dimensional words}\label{fig01}
\end{figure}

These 2-dimensional words are invariant with respect to translation by integer offsets, but not with respect to mirror or rotation. A word may have several disconnected components. Examples of words are presented in Fig.~\ref{fig01}: in (a) it is a rectangular word; in (b) it is a word of arbitrary shape having no holes and 1 component; in (c) there is a word with 1 hole and 2 components. 

\out{
$\begin{array}{c@{}c}
\genlet{a}	&\genlet{b}\vspace{-1mm}\\
\genlet{b}	&\genlet{c}
\end{array}$
$\begin{array}{c@{}c@{}c@{}c}
\genlet{a}	&			&\genlet{b}	&			\vspace{-1mm}\\
			&\genlet{c}	&\genlet{b}	&\genlet{a}	\vspace{-1mm}\\
			&\genlet{b}	&			&		\vspace{-1mm}\\
\end{array}$
}

\subsection{Local constraints: Tiles}

\begin{figure}
\begin{tabular}{c@{\hspace{1cm}}c}
\includegraphics[scale=0.3]{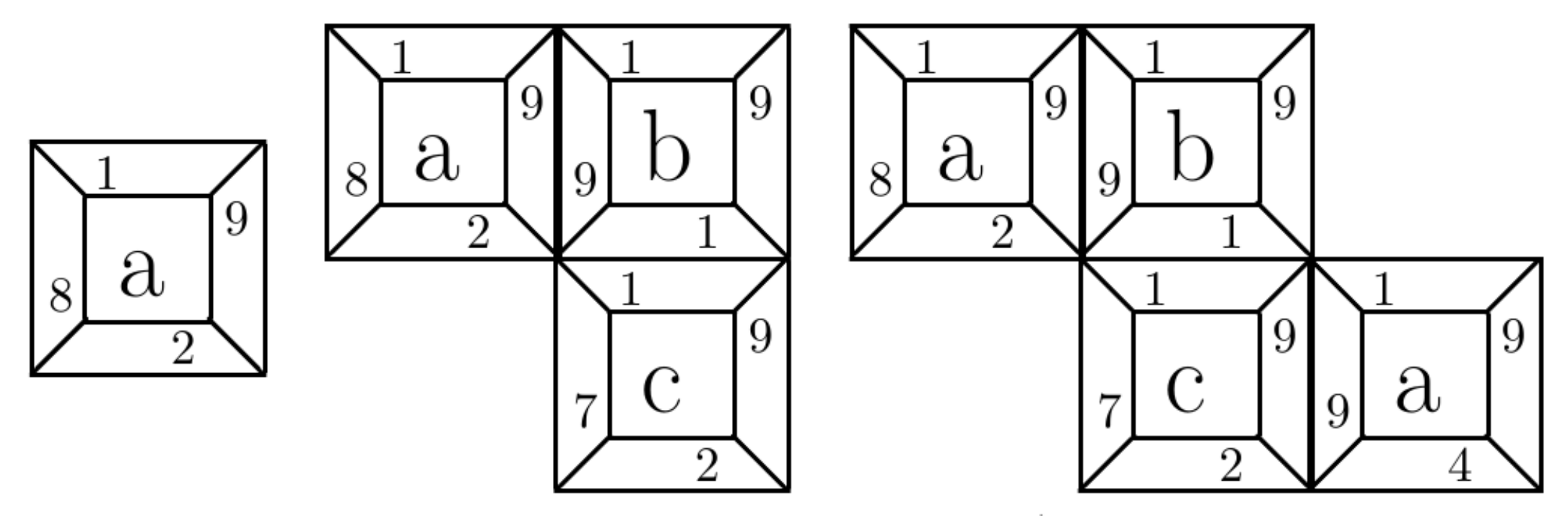}
& \includegraphics[scale=0.3]{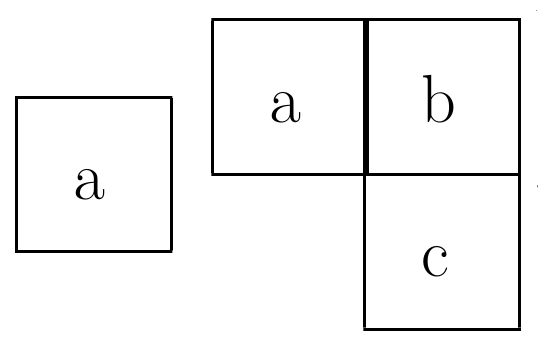}\\
(a)&(b)
\end{tabular}
\caption{Scenarios and accepted words}\label{fig02}
\end{figure}

A \textit{tile} is a letter enriched with additional information on each border. This information is represented abstractly as an element from a finite set\foo{Often, we use sets of numbers or sets of colors.} and is called a \textit{border label}. The role of border labels is to impose local glueing constraints on self-assembling tiles: two neighbouring cells, sharing a horizontal or a vertical border, should agree on the label on that border. A \textit{scenario} is similar to a 2-dimensional word, but: (1) each letter is replaced by a tile; and (2) horizontal or vertical neighbouring cells have the same label on the common border. Examples of tiles and scenarios are presented in Fig.~\ref{fig02}(a).

\out{
$\begin{array}{c}\gentile{a}{8}{1}{9}{2}\end{array}$
$\begin{array}{c@{}c}
\gentile{a}{8}{1}{9}{2}	&\gentile{b}{9}{1}{9}{1}	\vspace{-1mm}\\
						&\gentile{c}{7}{1}{9}{2}
\end{array}$
$\begin{array}{c@{}c@{}c}
\gentile{a}{8}{1}{9}{2}	&\gentile{b}{9}{1}{9}{1}	\vspace{-1mm}\\
						&\gentile{c}{7}{1}{9}{2}	&\gentile{a}{9}{1}{9}{4}
\end{array}$
}

A \textit{self-assembling tile system\foo{
Tiling is a popular research subject, see e.g. \cite{book-tiling87}. The model we are using here (with a finite label set for borders) was introduced in the context of interactive programming \cite{ste02,ste06a} using a different terminology. In that context, the concept is called \textit{finite interactive system (FIS)}. The vertical dimension represents \textit{time}, while the horizontal dimension represents \textit{space}. The labels on the north and south borders represent (abstract) \textit{memory states}, while the ones on the west and east borders represents (abstract) \textit{interaction classes}. The selected labels on the external borders are called \textit{initial} for west and north borders, and \textit{final} for east and south borders.}
} (shortly, \textit{SATS}) is defined by a finite set of tiles, together with a specification of what border labels are to be used on the west/north/east/south external borders. An \textit{accepting scenario} of a SATS $F$ is a scenario, obtained by self-assembling tiles form $F$, having the specified labels on the external borders. Finally, \textit{the 2-dimensional language recognized by a SATS $F$}, denoted $\cal{L}(F)$, is the set of 2-dimensional words obtained from the accepting scenarios of $F$, dropping the border labels.

\textit{Example:} An example of a SATS is $F$ defined by:\\
\hspace*{1cm}(1) tiles: \raisebox{-.3cm}{\includegraphics[scale=0.25]{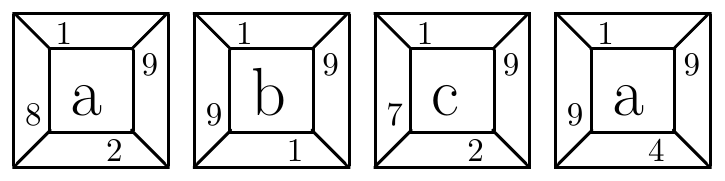}} and \\
\hspace*{1cm}(2) labels for external w/n/e/s borders: \{7,8\}/\{1\}/\{9\}/\{2\}.\\
All scenarios in Fig.~\ref{fig02}(a) are correct scenarios of $F$. The first two are accepting scenarios, while the last one is not (there is a label 4 on the south border). Dropping the border labels in the accepting scenarios in Fig.~\ref{fig02}(a) we get the recognized words in Fig.~\ref{fig02}(b).
\out{
$\begin{array}{c}\genlet{a}\end{array}$
$\begin{array}{c@{}c}
\genlet{a}	&\genlet{b}	\vspace{-1mm}\\
			&\genlet{c}
\end{array}$.
}

Tiles constrain the letters of horizontal and vertical neighbouring cells. They control the letters of those cells in a word laying in a \textit{horizontally-vertically connected component} (shortly \textit{hv-component})\foo{The words in Fig.~\ref{fig01}(a),(b),(c) have 1,3,7 hv-components, respectively.} of a word and do not affect separated components. Consequently, our focus below will be on the structure of hv-components of the words.

Projected on one dimension, this model produces classical \textit{finite automata}. For instance, this can be done by considering different labels for the west and the east borders of the tiles, inhibiting horizontal growth of the scenarios. Then each accepted hv-connected scenario is an 1-column scenario which may be seen as an acceptance witness of a usual 1-dimensional word by the corresponding finite automaton.  

It is not at all obvious how to define 2-dimensional word composition, extending usual 1-dimensional word concatenation. We start with the simpler definition of scenario composition. 
\svsp

\textbf{Scenario composition.}
For two scenarios $v$ and $w$, the \textit{scenario composite} $v.w$ consists of all valid scenarios resulting from putting $v$ and $w$ together, without overlapping. This actually means that, if $v$ and $w$ share some borders in a particular placement, then the labels on the shared borders should be the same. 
\svsp

\begin{figure}
\begin{tabular}{ccc}
\includegraphics[scale=0.2]{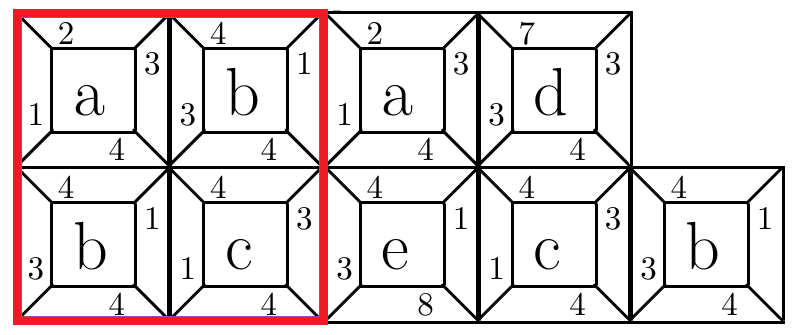}
& \includegraphics[scale=0.2]{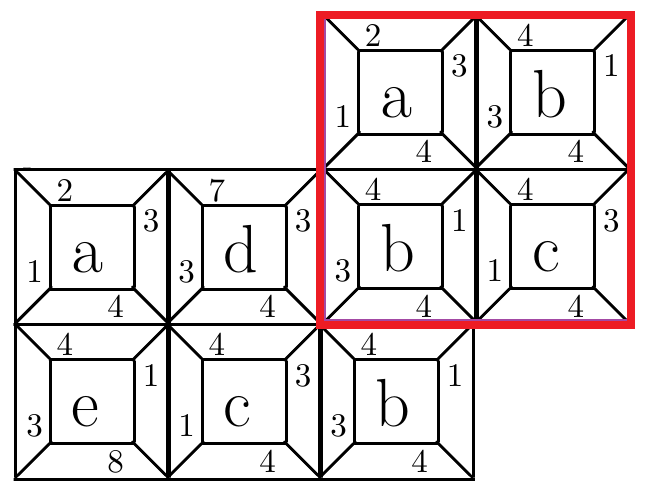}
& \includegraphics[scale=0.2]{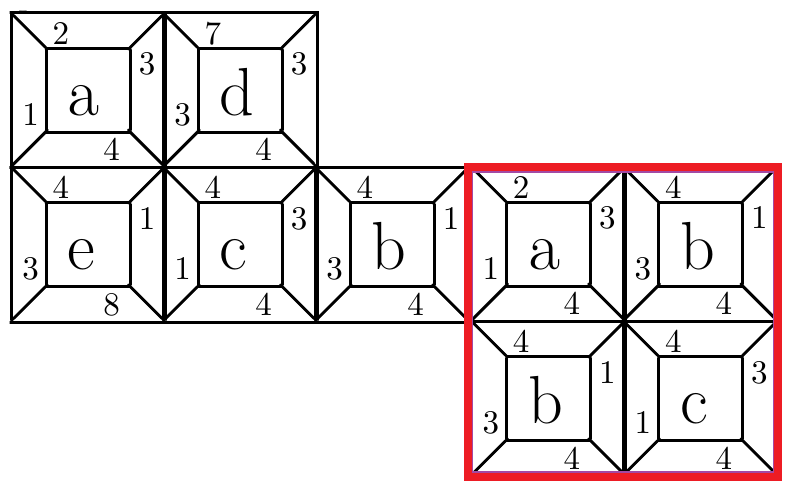}\\
(a)&(b)&(c)
\end{tabular}
\caption{Scenario composition}\label{fig03}
\end{figure}

\textit{Example:} We consider two scenarios:\\
\hspace*{1cm}$v= \raisebox{-1cm}{\includegraphics[scale=0.3]{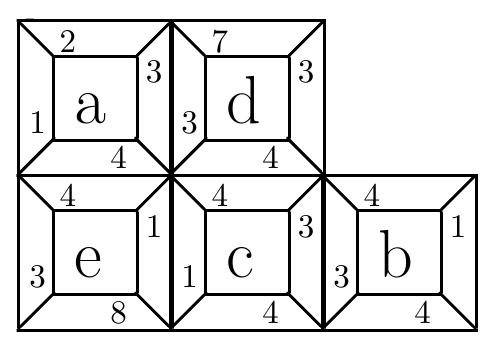}}
\hsp w= \raisebox{-1cm}{\includegraphics[scale=0.3]{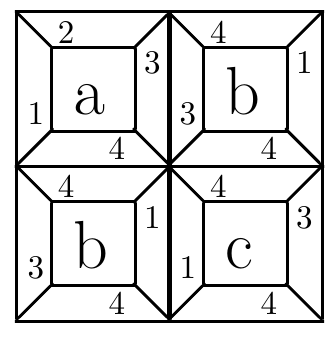}}$
\out{
$v=\begin{array}{c@{}c@{}c@{}c}
\gentile{a}{1}{2}{3}{4} & \gentile{d}{3}{7}{3}{4} &  & \vspace{-1mm}\\
\gentile{e}{3}{4}{1}{8} & \gentile{c}{1}{4}{3}{4} & \gentile{b}{3}{4}{1}{4} &  
\end{array}$
$w=\begin{array}{c@{}c}
\gentile{a}{1}{2}{3}{4}	&\gentile{b}{3}{4}{1}{4}\vspace{-1mm}\\
\gentile{b}{3}{4}{1}{4}	&\gentile{c}{1}{4}{3}{4}
\end{array}$
}
\\The composite $v.w$ has 3 results, sharing at least one cell border; they are presented in Fig.~\ref{fig03}.

\textbf{Word composition, a preliminary solution.}
The above scenario composition guides us towards what word composition should be able to achieve. 

One possibility is to mimic scenario composition using enriched alphabets and renaming. Notice that tiles may be codified using letters in appropriate enriched alphabets - one possibility in presented in the next paragraph. There are three strong criticisms to this solution. First, composition would depend on the underlying tile system and we want a word composition definition depending only on the words themselves. Second, the use of tiles (border labels) does not scale to large systems\foo{In one particular case, this is a well-known problem. When the tiling is on the vertical direction only, we get finite automata with labels for controlling sequential program executions. Large programs are difficult to cope with using ``go-to programs'' corresponding to finite automata. The well-known solution to avoid using labels in sequential programming is to use ``structured programming'' based on a particular class of regular expressions.}. Finally, and perhaps the most important reason, renaming does not behave properly in combination with intersection, a key operation used in classical regular expressions \cite{gi-re97,lmn98}; for instance, by renaming letters in an expression representing square words one gets an expression producing arbitrary rectangular words - see \cite{bps13}.

The drawbacks of this ``word composition via scenario composition'' is inherited by ``regular expressions'' including renaming operators \cite{gi-re97,lmn98}\foo{Renaming is also used in getting regular expressions for Petri nets \cite{ga-ra92} and timed automata \cite{acm02}.}. Indeed, with renaming we can mimic the border labels of the tiles as additional information in the cells letters; e.g., we can enrich letter $a$ to become $\ol{a}=_{def}(a,1,2,3,4)$, including into the cell letter the labels 1,2,3,4 of the side borders. Then, we can use this extra information on neighbouring cells to control the shape of the words and, finally, using a renaming, we can remove the additional information. Actually, with renaming the border labels are reintroduced in the model.

To conclude: \bi\item[]\textit{Scenario composition is based on a set of tiles, while word composition (and the associated regular expressions) should not have any underlying tile system behind.}\ei

\begin{figure}
\begin{tabular}{c}
\hspace{-.75cm}\includegraphics[scale=0.15]{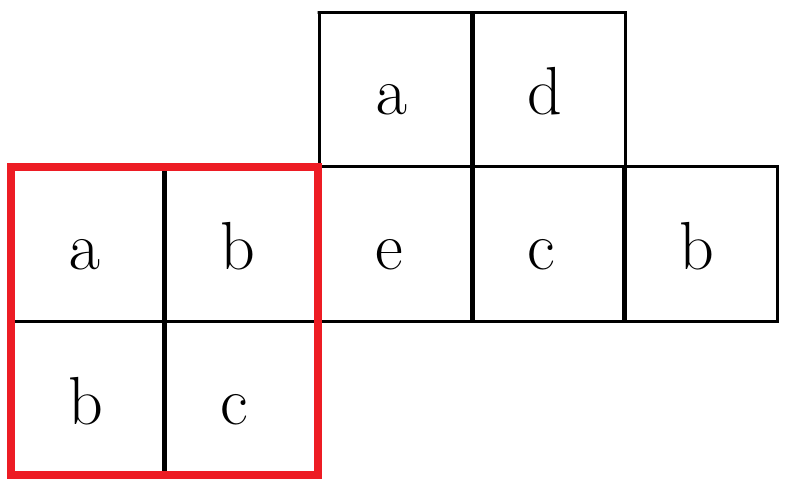}
\includegraphics[scale=0.15]{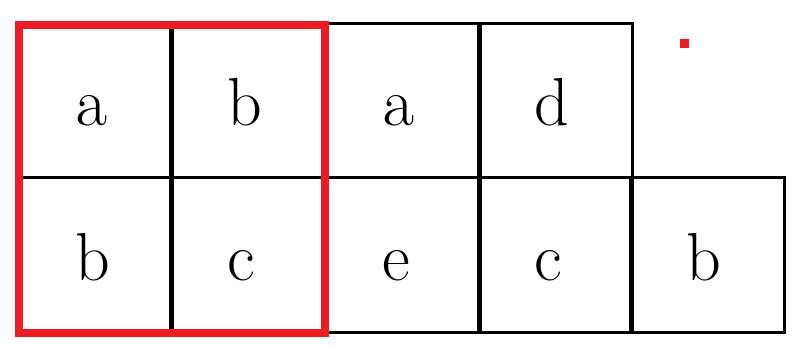}
\includegraphics[scale=0.15]{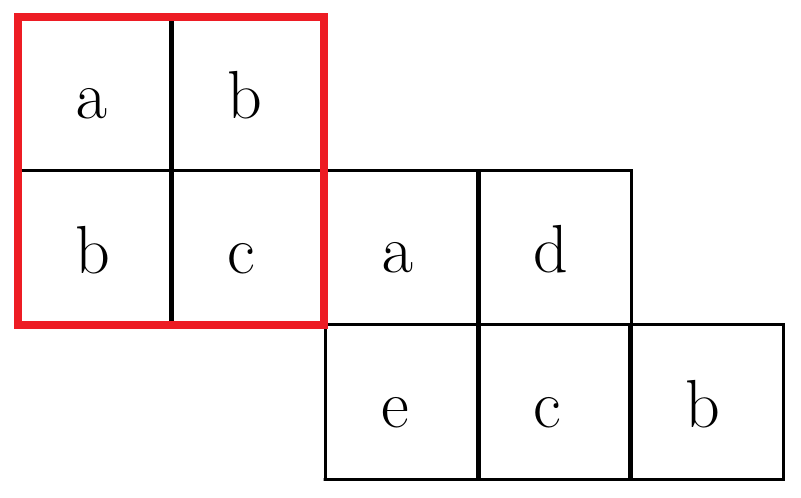}
\includegraphics[scale=0.15]{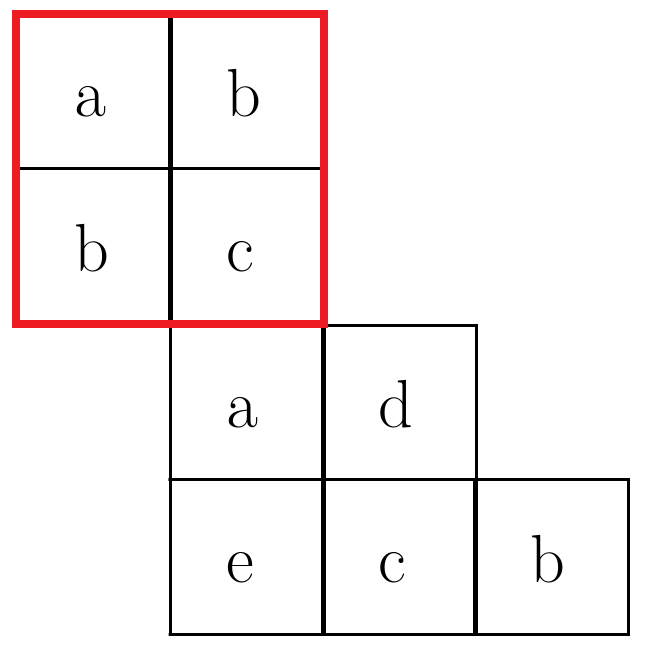} ...
\includegraphics[scale=0.15]{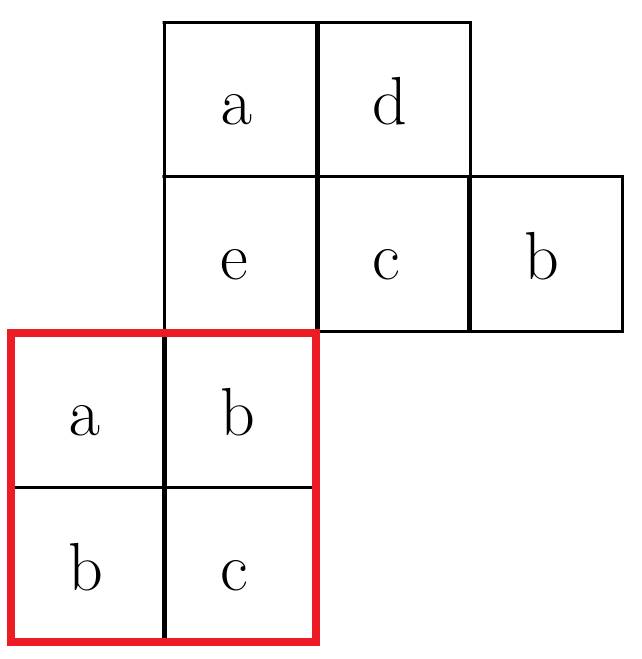}
\end{tabular}
\caption{Border agnostic word composition}\label{fig04}
\end{figure}

\textbf{Word composition.}
The solution to the problems identified in the previous paragraphs is to define a restricted composition using relevant information of the contours of the words \cite{bps13,ba-st14b,ba-st14a}, only. A few elements of interest on the contours are: 
\bi\item\textit{side borders} (w = west, i.e., the cell at the right of the side is inside the word; e = east; etc.), 
\item \textit{land corners} (nw = north-west corner seen from inside the word, i.e., the cell at the bottom-right of the point is inside the word, while the other 3 cells around are not inside the word; ne = north-east; etc.) and 
\item \textit{golf corners} (nw' = north-west corner seen from outside the word, i.e., the cell at the bottom-right of the point is outside the word, while those at the top-right and bottom-left are inside the word; ne' = north-east golf corner; etc.).\ei 

\begin{figure}
\begin{tabular}{ccc}
\includegraphics[scale=0.2]{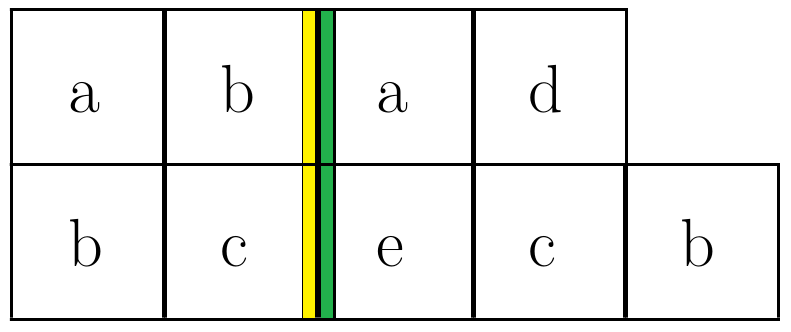}
& \includegraphics[scale=0.2]{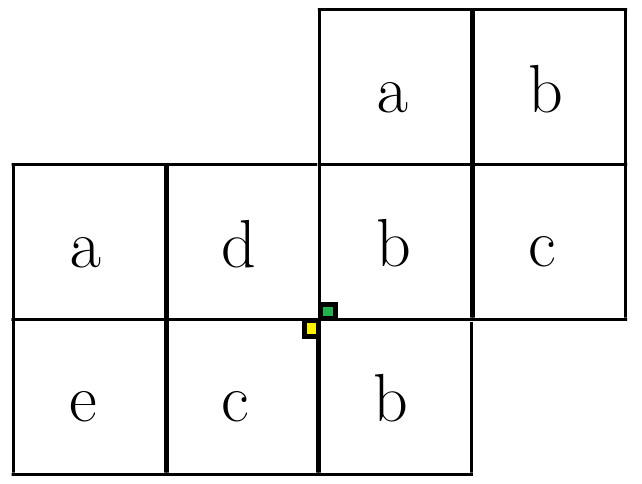}
& \includegraphics[scale=0.2]{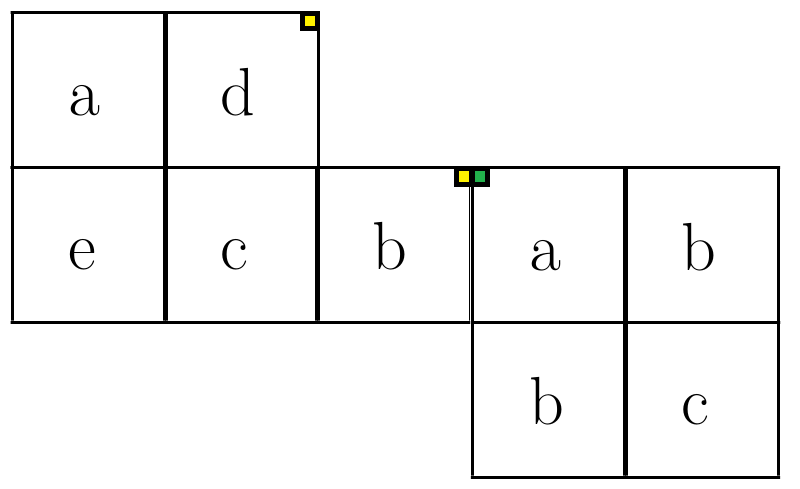}\\
(a)&(b)&(c)
\end{tabular}
\caption{Restricted word composition}\label{fig05}
\end{figure}

\textit{Example:} To get the words corresponding to the scenario composition in Fig.~\ref{fig03}, we can use compositions based on the following restrictions: \be
\item $v \texttt{(w=e)} w$ - the west border of $v$ is equal to the east border of $w$; this yields a result similar to the one presented in Fig.~\ref{fig03}(a);
\item $v \texttt{(sw'=sw)} w$ - the south-west golf corners of $v$ (there is only one) are identified with the south-west land corners of $w$ (there is only one); this restriction is good for Fig.~\ref{fig03}(b);
\item $v \texttt{(ne>nw)} w$ - the north-east corners of $v$ (there are two) includes the north-west corners of $w$ (there is only one); this is for Fig.~\ref{fig03}(c).\ee
These restricted compositions are illustrated in Fig.~\ref{fig05}.
\svsp 

\textit{Remark:}
Corner composition should be used with care. For instance, we can constrains the order of elements on a diagonal as in $(a \texttt{(se=nw)} b) \texttt{(se=nw)} c$, which is not possible using tiles alone, as they only constrain horizontally-vertically connected cells. However, it can be done if we have cells near diagonal to ensure hv-connections between the cells on the diagonal. 
\vsp 

\begin{figure}
\begin{tabular}{c}
\hspace*{0.cm}\includegraphics[scale=.5]{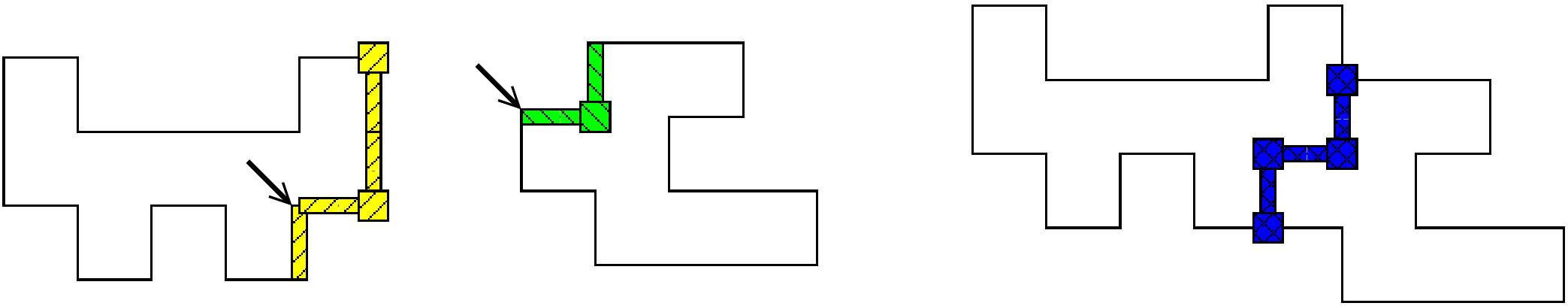}\\
\hspace*{0.8cm}(a) $v$\hspace{2.2cm}(b) $w$\hspace{2.2cm}(c) $v\ R(Y,G,B)\ w$\hspace*{0.8cm}
\end{tabular}
\caption{General restricted composition}\label{fig06}
\end{figure}

\textit{General restricted word composition:}
A general \textit{restricted composition} is defined as follows. Suppose we are given: 
\be\item Two words $v,w$; a subset $Y$ of elements of the contour of $v$ (emphasized elements on the contour of $v$ as in Fig.~\ref{fig06}); a subset $G$ of elements of the contour of $w$ (emphasized elements on the contour of $w$); the subset $B$ of actual contact elements after composing $v$ with $w$ via the points indicated by the little arrows (the emphasized elements in the composed $R(Y,G,B)$ picture in Fig.~\ref{fig06}).
\item a relation $R(Y,G,B)$ between the above 3 subsets\foo{For the example in Fig.~\ref{fig06}, a relation $R$ making the restricted composition valid may be: $G\subseteq Y \wedge G\subseteq B$ (after composition, all the emphasized elements on the contour of $w$ are on the common border and included in the set of emphasized elements of $v$).}.
\ee
The resulted restricted composition is denotes by $v\ R(Y,G,B)\ w$.
\vsp

\out{
A further attribute for restricted composition is the distinction between the normal formula satisfiability and another one, called \textit{strict interpretation}. In the latter interpretation the formula characterize the whole set of border connections between the words.
}

\subsection{Global constraints: Regular expressions}

\textbf{Regular expressions.} A new approach for defining classes of two-dimensional regular expressions has been introduced in \cite{bps13,ba-st14b,ba-st14a}. It is based on words with arbitrary shapes and classes of restricted composition operators. 

The basic class \textit{n2RE} uses compositions and iterated compositions corresponding to the following restrictions (see \cite{ba-st14a} for more details and examples):\be\item the selected elements of the words contours are: \textit{side borders}, \textit{land corners}, and \textit{golf corners};
\item the atomic comparison operators are: \textit{equal-to} `\texttt{=}', \textit{included-in} `\texttt{<}', \textit{non-empty intersection} `\texttt{\#}';
\item the general comparison formulas are boolean formulas built up from the atomic formulas defined in 2.
\ee

An enriched class \textit{x2RE} \cite{ba-st14a} is obtained adding ``extreme cells'' glueing control. A cell is \textit{extreme} in a word if it has at most one neighbouring cell in that word, considering all vertical,  horizontal, and diagonal directions. The restricted composition may use elements of interest of the contours belonging to extreme cells. They are denoted by prefixing the normal \textit{n2RE} restrictions with an `\texttt{x}'; e.g., \texttt{xw}, \texttt{xse}, \texttt{xnw'}, etc. For instance, $v \texttt{(e>xw)} w$ is true if the west borders of the extreme cells in $w$ are included in the east borders of $v$.
 
\subsection{Systems of recursive equations}

A system of recursive equations is defined using variables representing sets of (two-dimensional) words and regular expressions. Formally, a \textit{system of recursive equations} is represented by:
$$\left\{\begin{array}{rcl}
X_1&=&\sum_{i_1=1,k_1}E_{1i_1}(X_1,\dots,X_n)\\
&\dots&\\
X_n&=&\sum_{i_n=1,k_n}E_{ni_n}(X_1,\dots,X_n)\\
\end{array}\right.$$
where $X_i$ are variables (denoting sets of words) and $E_{ij}$ are regular expressions over a given alphabet, extended with occurrences of variables $X_i$. 
\svsp

\textit{Comments:} A basic operation here is the substitution of 2-dimensional word languages for the variables $X_{i}$ used in these equations. A similar operation, but restricted to rectangular words, has been used in \cite{ccpp,cp} to define tile grammars. Our formalism, using arbitrary shape words, may be seen as a generalization of this mechanism. 
\svsp

%\raisebox{.7cm}{\tdwbis{small}{xaaax\tdret bxaxc\tdret bbxcc\tdret bxdxc\tdret xdddx}}

\begin{figure}
\begin{tabular}{c}
\out{
\includegraphics[scale=.20]{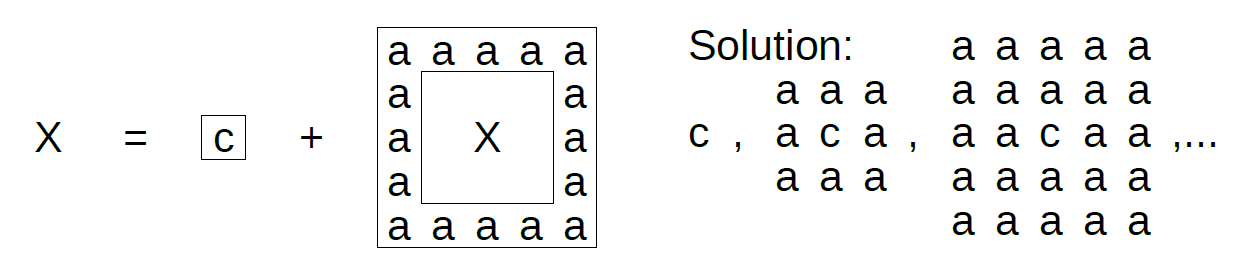}\\
\includegraphics[scale=.20]{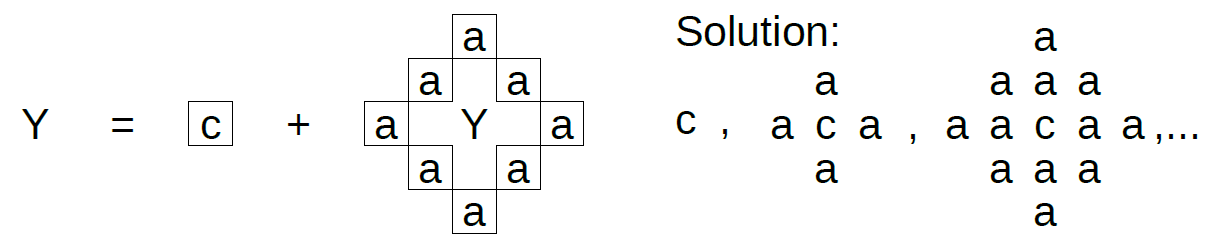}
}
\includegraphics[scale=.20]{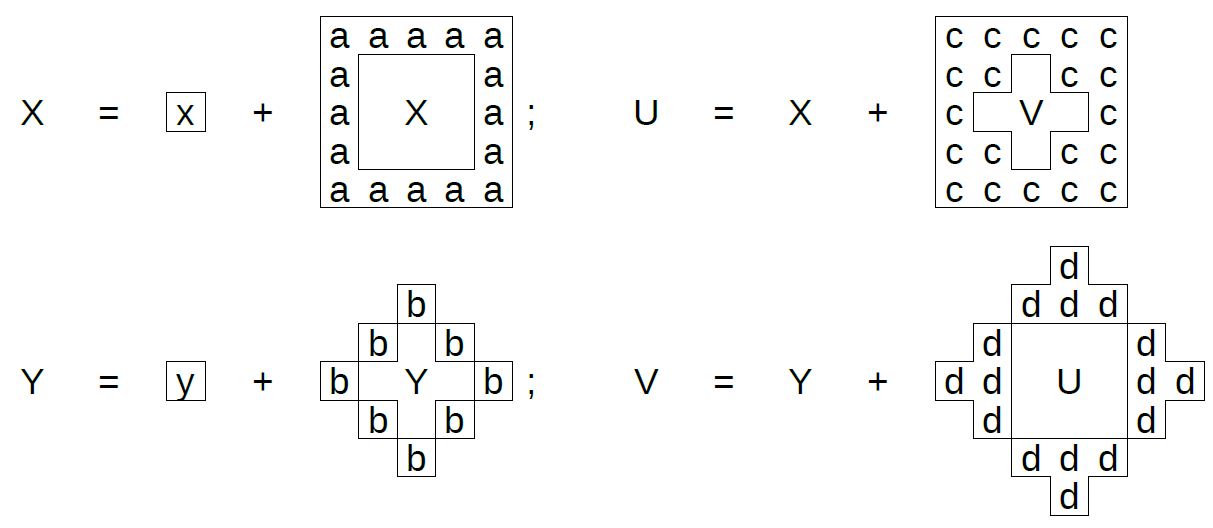}
\end{tabular}
\caption{Recursive equations}\label{fig21}
\end{figure}

\textit{Example 1:} The language consisting of square words filled with $a$, except for the center which contains $x$, may be represented by the system/equation (this equation is illustrated by the equation for $X$ in Fig.~\ref{fig21}):
\bi\item[(*)] $X= x + E(X)$\ei 
where: 
\bi\item[]$\begin{array}{r@{\ }c@{\ }l}
E_r&=& (_3(_1a \texttt{*(e=w)})_1\texttt{(se=ne)}(_2a\texttt{*(s=n)})_2)_3\\
&& (_5\texttt{(sw=ne)})_5\ (_4(_1a \texttt{*(e=w)})_1\texttt{(nw=sw)}(_2a\texttt{*(s=n)})_2)_4\\
E_{rect}&=&(E_r (_6\texttt{(nw>ne)\&(nw>sw)})_6 a)\ (_7\texttt{(se>ne)\&(se>sw)})_7\ a\\
E(X)&=& X (_8\texttt{(n<s)\&(e<w)\&(s<n)\&(w<e)})_8 E_{rect}
\end{array}$\ei
The expression within the parentheses with index 1 in $E_{r}$ generates horizontal bars of $a$'s. The one in 2 produces vertical bars. The restrictions in 3 and 4 orthogonally glue the bars in corners, yielding `\raisebox{-.1cm}{$\urcorner$}' and `$\llcorner$' shapes, respectively. Finally, 5 constraints these two corners to produce a rectangle, without the \texttt{nw} and \texttt{se} corners. Finally, the constraints 6 and 7 in $E_{rect}$ fill in $a$'s in these corners. For $E(X)$, the restriction 8 requires $X$ to have all borders included in those of $E_{rect}$, hence $X$ has to be a rectangle itself and to fill the whole interior part of the rectangle specified by $E_{rect}$. Finally, the recursive procedure (*) starts with a square $x$, so we get precisely the required square words.

\textit{Example 2:} The example above can be adapted to produce square diamonds of $b$'s, having $y$ in the center ($Y$ in Fig.~\ref{fig21}). Indeed, instead of the restrictions $\texttt{(e=w)}$ and $\texttt{(s=n)}$, used in 1 and 2, one can use $\texttt{(se=nw)}$ and $\texttt{(sw=ne)}$, respectively, to produce diagonal bars. To produces corners, as in 3 and 4, is slightly more complicated: use extreme cells to locate the corners in the heads of the bars to be connected. The remaining part is similar. As a last remark, notice that the resulting expression is in \textit{x2RE}, not in \textit{n2RE}. 

\textit{Example 3:} Finally, $U$ and $V$ in Fig.~\ref{fig21} describe a mutually recursive construction built on top of the languages in the previous examples. 

\section{Structural characterisation for self-assembling tiles}

In this section we present the proposed research plan.

\subsection{From local to global constraints}

A main question in understanding the tiling procedure is the connection between: (1) the local representation using tiles, scenarios, and labels on borders; and (2) the global representation making use of regular expressions and systems of recursive equations. 

Here, we are particularly interested in one direction:\snvsp 
\bi\item[Q1:] \textit{Is is possible to get representations by systems of recursive equations for all tile systems?}\snvsp
\item[Q2:] \textit{What is a minimal set of regular expressions expressive enough to represent tile systems?}\snvsp
\item[Q3:] \textit{Can we find a kind of ``normal form representation'' for these systems of recursive equations?}\snvsp
\ei And finally,\snvsp\bi 
\item[Q4:] \textit{Is it possible to develop a (correct and complete) algebraic approach for modelling tile systems?}\snvsp
\ei

\subsection{Languages generated by two-colors border tiles}

A minimal, non-trivial SATS should have at least 2 labels for each vertical and horizontal dimension. Up to a bijective representation, the tiles of a SATS using 2 distinct labels for each vertical and horizontal dimension can be seen as elements in the following set
\bi\item[]\hspace*{-.7cm}\tta\ \ttb\ \ttc\ \ttd\ \tte\ \ttf\ \ttg\ \tth\ \tti\ \ttj\ \ttk\ \ttl\ \ttm\ \ttn\ \tto\ \ttp. \ei
In this representation, the labels for the vertical and the horizontal borders are denoted by 0/1. 
\out{Equivalently, we use colors: for  vertical borders, green/yellow are used instead of 0/1, while for horizontal borders, blue/pink replace 0/1.} 
The letter associated to a tile is the hexadecimal number obtained from the binary representation of the sequence of its west-north-east-south 0/1 digits, in this order; for instance, the label of the tile \ttl\ is the number represented by $1011$, hence $b$.
\svsp

\textbf{Notation:} \textit{(1) Tiles:} If $A=\{t_1,\dots,t_k\}$ is a subset of $\{0,1,\dots,f\}$, then a SATS defined by the tiles in $A$ is denoted by $Ft_1t_2\dots t_k$. To have an example, $F02ac$ consists of tiles \tta\ \ttc\ \ttk\ \ttm. 

\textit{(2) External labels:} A SATS also uses labels for the external borders of its accepting scenarios. The recursive specifications bellow will use variables $X_{wnes}$ denoting the set of words recognised by scenarios with $w/n/e/s$ on their west/north/east/south borders, respectively. Here, for simplicity, \textit{we add the restriction} to have \textit{only one} label for each west/north/east/south external border. Then, $Ft_1t_2\dots t_k$ completely defines a SATS by specifying the labels used for the external borders. There are at most 16 possibilities, each one denoted by a hexadecimal number representing the sequence of the west-north-east-south 0/1 labels used for the external borders. 

\textit{(3) Final notation:} To conclude, the SATSs to be investigated are represented as $Ft_1t_2\dots t_k.z$, where $t_1, t_2, \dots, t_k$ are the tiles and the binary digits of $z$ specify the labels used for the external borders. 

As an example, $F02ac.c$ consists of: (1) tiles \tta\ \ttc\ \ttk\ \ttm\ and (2) labels for external borders: 1 on west, 1 on north, 0 on east, and 0 on south.
\vsp 

\textbf{The research programme:} The goal of the proposed research programme is to:\snvsp 
\be
\item[P1] \textit{Find representations by systems of recursive equations for the languages generated by all SATSs $Ft_1\dots t_k.z$ (there are 1048576 cases to consider - $2^{16}$ subsets of tiles and $2^4$ combinations of border labels for each subset);}\snvsp
\item[P2] \textit{Extend these representations to SATSs with any number of labels for borders (and any number of labels for external borders).}\snvsp
\ee

\subsection{A case study: $F02ac.c$}

To start the analysis, note that we can construct any shape of \tta\ and \ttk, and vertical bars of \ttc. The last tile \ttm\ can be composed with itself, in a connected word, only along the diagonal. There are 4 possible direction changes for \ttm, denoted by letters $X,Y,Z,T$. A direction change in a hv-connected component is possible if: (1) one can insert one tile in interior, as in $X3$; or (2) one can insert two tiles on exterior areas as in $X1$ and $X2$. Three interior direction changes via $Y3,Z3,T3$ are not possible; the last one, $X3$, may be filled with \ttc. On the other hand, via exterior connections only the combination $Z1\& Z2$ is possible using, say, \ttk\ and \tta. 

$\begin{array}{c@{}c@{}c@{}c@{}c@{}c}
&X1&\ttm&X2&&\\
&\ttm&X3&\ttm&&\\
&&&&\ttm&Y1\\
T1&\ttm&&&Y3&\ttm\\
\ttm&T3&&&\ttm&Y2\\
T2&\ttm&Z3&\ttm&&\\
&Z1&\ttm&Z2&&
\end{array}$
\hsp
$\begin{array}{c@{}c@{}c@{}c@{}c@{}c@{}c@{}c@{}c}
&&&&\ttm&&&&\\
&&&\ttm&\ttc&\ttm&&&\\
&&\ttm&&&&\ttm&&\\
&\ttm&&&U&&&\ttm&\\
\ttm&&&&&&&&\ttm
\end{array}$

To fill in the area $U$ we need an horizontal passing from border 0 to 1, and this may be done by horizontal words from the expressions \texttt{0$^*$2a$^*$}.

$\begin{array}{c@{}c@{}c@{}c@{}c@{}c@{}c@{}c@{}c@{}c@{}c@{}c}
&&&&&&\ttm&&&&\\
&&&&&\ttm&\ttc&\ttm&&&\\
&&&&\ttm&\ttc&\ttk&\ttk&\ttm&&\\
&&&\ttm&\tta&\tta&\ttc&\ttk&\ttk&\ttm&\\
&&\ttm&\tta&\tta&\tta&\tta&\tta&\ttc&\ttk&\ttm\\
&\ttm&\tta&\tta&\tta&&\ttk&\ttk&\ttk&\ttk&\ttk&\ttm\\
\ttm&\tta&\tta&
\end{array}$
\vsp

The north selected label is 1, hence the top of each column should start with an \texttt{c}. Hence, in a first approximation, a hv-connected component of an accepted word should have the \textit{hat-form} above. 

The general format of a hv-component is slightly more complicated: two hat-forms, as before, can be connected via the cells of their extreme bottom legs as in  
$\begin{array}{c@{}c@{}c@{}c@{}c@{}c@{}c@{}c@{}c@{}c@{}c@{}c}
\ttm&&&&\ttm&\\
\ttk&\ttm&&\ttm&\ttc&\ttm\\
&\ttk&\ttm&\tta&&
\end{array}$.
\svsp

To get a recursive specification for the hat-form we proceed as follows:\snvsp\be
\item construct a shape $C1$ for \texttt{c}'s as two diagonals connected on the top cells;\snvsp
\item construct horizontal bars with \texttt{0}'s on the left, \texttt{a}'s on the right, and separated by a \texttt{2};\snvsp
\item iteratively fill with these bars the shape $C1$, requiring to completely connect their west, north, and east borders; let $C2$ the the resulting word;\snvsp
\item finally, iteratively connect $C2$ with horizontal bars of \texttt{0}'s on west-north and horizontal bars of \texttt{a}'s on north-est, completely connecting their west-north and north-east borders, respectively.\snvsp
\ee
It is not difficult to see that this procedure may be formalized by a system of recursive equation using expressions in $n2RE$. One possibility is $X11$ defined by the following system of recursive equations:

\begin{figure}
\begin{tabular}{c}
\includegraphics[scale=0.25]{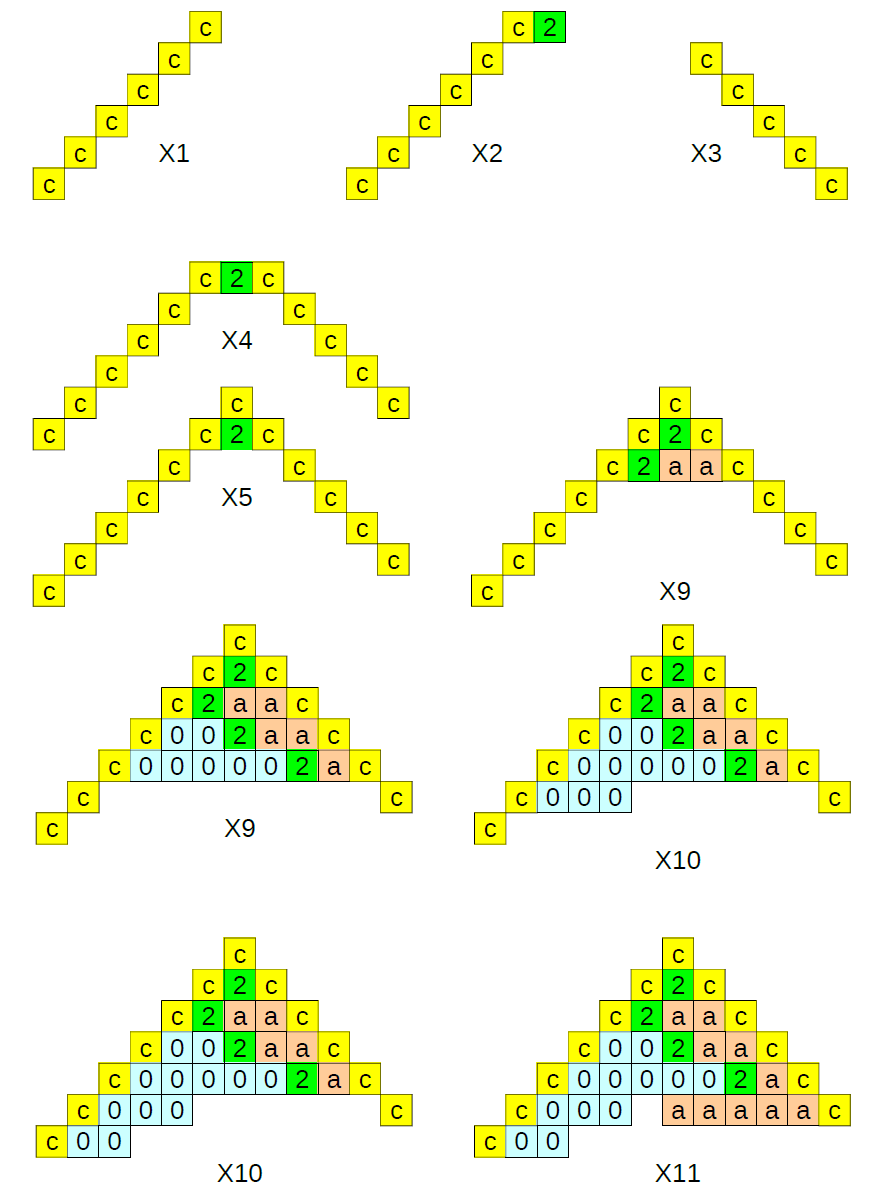}
%\hspace*{-.3cm}\includegraphics[scale=0.19]{chapters/chapterGS/fig/spec-rec3}
\end{tabular}
\caption{Recursive specifications}\label{fig54}
\end{figure}

\bi\item[]$
X1 = c + c\ \texttt{(sw=ne)}\ X1\\
X2 = X1\ \texttt{(ne=nw)}\ 2\\
X3 = c + c\ \texttt{(se=nw)}\ X3\\
X4 = X2\ \texttt{(ne=nw)}\ X3\\
X5 = c\ \texttt{(s<n)\&!(sw\#nw)\&!(se\#ne)}\ X4\\
X6 = ((0\ \texttt{*(e=w)})\ \texttt{(e=w)}\ 2) \texttt{(e=w)}\ (a\ \texttt{*(e=w)})\\
X7 = (0\ \texttt{*(e=w))}\\
X8 = (a\ \texttt{*(e=w)})\\
X9 = X5 + X6\ \texttt{(n<s)\&(w<e)\&(e<w)}\ X9\\
X10 = X9 + X7\ \texttt{(n<s)\&(w<e)}\ X10\\
X11 = X10+X8\ \texttt{(n<s)\&(e<w)}\ X11
$\ei
The patterns defined by $X1,\dots,X11$ are illustrated in Fig.~\ref{fig54}.

For the general format, consisting in connected hat-forms, one can use an $x2RE$ variation of the previous system. Use an additional $X5'$ defined by (\texttt{!x} means non-extreme)
\bi\item[]$X5'=X5 + X5 \texttt{((xne<xsw)\&!((!x)nw\#sw))} X5'$\ei
and change $X9$ to 
\bi\item[]$X9 = X5' + \dots$\ei

\section{Interactive programs}

A classical slogan states that ``program = control + data''. The control part for simple sequential programs is provided by finite automata. We can extend this saying that ``distributed program = (control~\&~interaction) + data''. The control~\&~interaction part may be specified by SATSs. In this section we briefly show how data can be added to SATSs to get completely specified distributed programs. Actually, one can use either SATSs or regular expressions for specifying the control~\&~interaction part of a distributed program.

\subsection{Words and traces, in two dimensions}

Recently, 2-dimensional languages have been used to study \textit{parallel, interactive, distributed systems} \cite{ste06a,dr-st08b,dr-st08a,ba-st14a,dpss14}; we simply call them \textit{interactive systems}. In these studies, the approach is less syntactical considering words up to a \textit{graph-isomorphism equivalence} \cite{dlpss12,dpss14}. This means, one uses two types of letters: \textit{connectors} in a set $C$ and (uninterpreted) \textit{statements} in a set $X$. Then, two words over $C\cup X$ are considered equivalent if they have the same occurrences of letters in $X$ connected in the same way via the elements in $C$. In other words, the placement of the letters in $X$ in the cells of a word does not matter, as long as elements in $C$ are used to ensure the connecting structure between the elements in $X$ is the same. 

The model of SATSs is not universal. To compare this class  with 1-dimensional languages, one can consider an 1-dimensional projection, for instance considering only the top row of the accepted rectangular 2-dimensional words. This way one gets a class of usual languages; by a theorem in \cite{la-si97}, this class coincides with the class of context-sensitive languages. A universal class can be easily obtained putting more information around the letters of the words. Scenarios for real computations, as the ones used in rv-systems \cite{ste06a} or in Agapia programming \cite{dr-st08a,pss07}, have complex spatial and temporal data attached to the cells. They are universal.

In most studies on 2-dimensional languages there is no distinction between the vertical and the horizontal dimension. The interpretation used in interactive systems is somehow different, considering the vertical dimension to represent the \textit{temporal evolution} of the system, while the horizontal dimension takes into account the \textit{spatial aspects} of the system. This way, a natural notion of \textit{2-dimensional trace} occurs: a column represents a process run (in an interacting environment), while a row describes a ``transaction'', i.e., a chain of process interactions via message passing (in a state dependent processing environment). 

A notion of \textit{trace-based refinement} for (structured) interactive systems has been recently presented in \cite{dpss14}. \textit{Traces} represent running scenarios modulo graph-isomorphism, projected on classes and states. In this interpretation one focuses on data stored in or flowing through the cells of the interactive system. 

\subsection{Interactive modules and programs}

In a SATS, a tile \includegraphics[scale=.07]{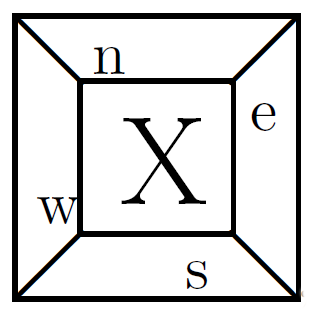}\out{$\gentile{X}{w}{n}{e}{s}$}, linearly represented as $X:\mytup{w}{n}\ra \mytup{e}{s}$, uses elements from an abstract finite set to label the borders. These labels are used for defining the control and the interaction used in interactive systems. In concrete, executable interactive systems the control/interaction labels are enriched with data. The data on the north and south borders are represented as spatial data implemented on memory, while the data on the west and east borders are temporal data implemented on streams. The former spatial data represent the states of the interactive processes, while the latter temporal data represents the messages flowing between processes. 

An \textit{interactive module} is a cell with: (1) control/interaction labels and temporal/spatial data on its borders and (2) a specification of a relation between border data. The relation describing the module functionality may be specified with a program, or in another way. 
\svsp

\textbf{Example: A communication protocol.}  We present a distributed program for a communication protocol between two parties using a channel which, due to time constrains, may decide to drop some data. During the first attempt of sending data, the sender keeps all the messages, while the receiver keeps two sets: one with received messages and one with the indices of missing ones. Then, the receiver sends indices of missing elements one by one to the sender waiting to receive those missing elements. 

We use the convention that $^\frown$ separate data on west or east borders coming from different modules, while $^\smile$ is similarly used, but for north or south borders. The overall behaviour of the protocol, described by the scenario below, is:
$$Protocol:\mytup{a^\frown b^\frown c}{0,\emptyset}\ra\mytup{a^\frown b^\frown c}{\emptyset}$$
First we give the specification of the modules, then we present the scenario.
\vsp\\\textbf{Modules:}
 
\texttt{Send-and-Keep:} 
\\$SK:\mytup{x}{i,Y}\ra\mytup{(i,x)}{i+1,Y\cup\{(i,x)\}}$;

\texttt{Communicate-Yes/No:}
\\$CY:\mytup{(i,x)}{\ }\ra\mytup{(i,x)}{\ }$; \hsp
$CN:\mytup{(i,x)}{\ }\ra\mytup{(i,?)}{\ }$;

\texttt{Receive-and-Keep:}
\\$RK:\mytup{(i,x)}{U,V}\ra\mytup{\ }{U,V\cup\{(i,x)\}}$; \
$RK:\mytup{(i,?)}{U,V}\ra\mytup{\ }{U\cup\{i\},V}$

\texttt{Send-first-bunch-End:}
\\$SEnd:\mytup{\ }{i,U}\ra\mytup{(i,end)}{U}$;

\texttt{Receive-first-bunch-End:}
\\$REnd:\mytup{(n,end)}{U,V}\ra\mytup{i}{U\setminus\{i\},V}$, for $i\in U$\\
$REnd:\mytup{(n,end)}{\emptyset,V}\ra\mytup{OK}{V}$

\texttt{Receive-and-Keep-at-Request:}
\\$RKR:\mytup{(i,x)}{U,V}\ra\mytup{j}{U\setminus\{i\},V\cup\{(i,x)\}}$, for $i\in U$ and $j\in U\setminus\{i\}$;\\ 
$RKR:\mytup{(i,x)}{U,V}\ra\mytup{OK}{V\cup\{(i,x)\}}$, if $U=\{i\}$;\\
$RKR:\mytup{(i,?)}{U,V}\ra\mytup{i}{U,V}$, for $i\in U$;

\begin{figure}
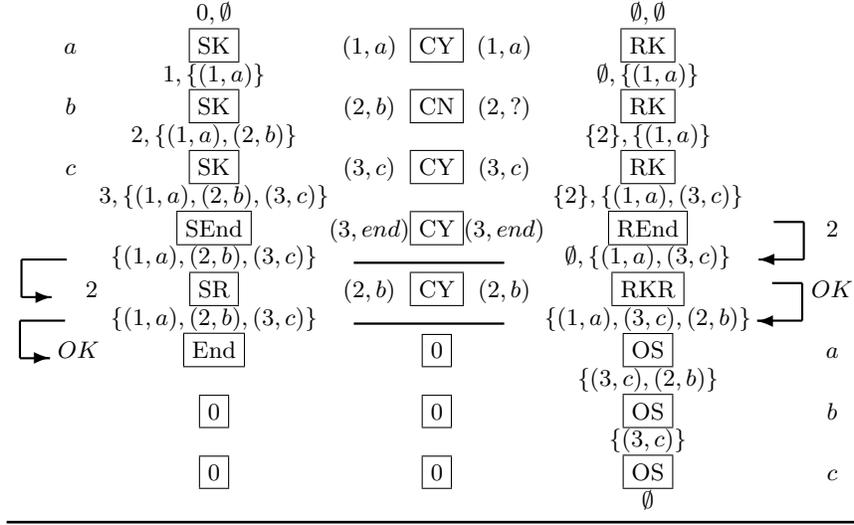

\begin{tabular}{c}
\begin{small}$\begin{array}{c@{}c@{}c@{}c@{}c@{}c@{\hspace{.8cm}}c}
&0,\emptyset &&&&\emptyset,\emptyset&\\
a&\framebox{SK}&(1,a)&\framebox{CY}&(1,a)&\framebox{RK}&\\
&1,\{(1,a)\}&&&&\emptyset,\{(1,a)\}\\
b&\framebox{SK}&(2,b)&\framebox{CN}&(2,?)&\framebox{RK}&\\
&2,\{(1,a),(2,b)\}&&&&\{2\},\{(1,a)\}\\
c&\framebox{SK}&(3,c)&\framebox{CY}&(3,c)&\framebox{RK}&\\
&3,\{(1,a),(2,b),(3,c)\}&&&&\{2\},\{(1,a),(3,c)\}\\
&\framebox{SEnd}&(3,end)&\framebox{CY}&(3,end)&\framebox{REnd}&2\\
&\{(1,a),(2,b),(3,c)\}&&&\hspace*{-2cm}\rule{2.cm}{.03cm}&\emptyset,\{(1,a),(3,c)\}\\
\baz{.5}{10.4}\hspace{.45cm} 2&\framebox{SR}&(2,b)&\framebox{CY}&(2,b)&\framebox{RKR}&OK\\
&\{(1,a),(2,b),(3,c)\}&&&\hspace*{-2cm}\rule{2.cm}{.03cm}&\{(1,a),(3,c),(2,b)\}\\
\baz{.5}{10.4}\hspace{.1cm}
OK&\framebox{End}&&\framebox{0}&&\framebox{OS}&a\\
&&&&&\{(3,c),(2,b)\}\\
&\framebox{0}&&\framebox{0}&&\framebox{OS}&b\\
&&&&&\{(3,c)\}\\
&\framebox{0}&&\framebox{0}&&\framebox{OS}&c\\
&&&&&\emptyset\\
\end{array}$\end{small}
\end{tabular}
\caption{A scenario for the communication protocol example}\label{fig-prot}
\end{figure}

\texttt{Output-Stream:}
\\$OS:\mytup{\ }{V}\ra\mytup{x}{V\setminus\{(i,x)\}}$, if $(i,x)\in V$ and $i=min\{j\co (j,y)\in V\}$;
\svsp\\\textbf{Scenario:}
An example of running scenario\foo{The drawn ``back-arrows'' represent a short notation for diagonal composition \cite{dr-st08a,dr-st08b}. The `0' cells may be omitted if we do not stick to have rectangular words/scenarios.} is presented in Fig.~\ref{fig-prot}.
\svsp

Interactive programs may be introduced on top of interactive modules in two ways: (1) either in an unstructured way using tiles with labels, or (2) in a structured way using (particular) regular 2-dimensional expressions. The rv-programs in \cite{ste04,ste06a} use the first option. On the other hand, Agapia structured interactive programming \cite{dr-st08a,pss07} is based on the second possibility, using rectangular 2-dimensional words/scenarios and 3 particular composition operators, together with their iterated versions: \textit{horizontal}, \textit{vertical}, and \textit{diagonal composition}.

\subsection{Refinement of structured interactive programs}

The presented formalism, based on self-assembling interactive modules, can be used as a basis for a refinement-based design. The approach consists in the following steps:\snvsp\bd
\item[Basic step:] A starting simple specification $S0$ may be introduced using independent specifications for process behaviours (columns) and for transactions (chains of interactions used in rows). In this step finite automata are used. \snvsp
\item[Matching control and interaction:] This step refines $S0$ to a specification $S1$ using  scenarios with a finite number of border labels (i.e., memory states and interaction classes). For $S1$ use a SATS\foo{Recall that what we call here a ``\textit{SATS}'' was introduced and used under the name ``\textit{finite interactive system}'' in this context; see \cite{ste04,ste06a} and the subsequent papers.} for control and interaction. Refinement correctness here requires the rows and the columns of $S1$ satisfy specification $S0$.\snvsp
\item[Adding data:] In this step, a new specification $S2$ is obtained by adding to the border labels concrete data for memory states and interaction messages around each scenario cell. This way one gets running scenarios describing concrete computations. The correctness criterion in this step is easy to formulate: by dropping the additional data and keeping the labels only, one must get simple patterns of control and interaction satisfying $S1$.\snvsp
\item[Iterated data and computation refinement:] Iterated refinement of data and computation, preserving trace semantics.\snvsp
\ed

This trace-based refinement design is presented in an abstract way here, no programming language being involved. Actually, it was introduced in combination with the programming language Agapia for describing running scenarios in structured interactive programs \cite{dpss14}. An open problem is to lift this trace-based refinement definition to a refinement definition on Agapia programs themselves.

\section{Conclusions}

Sequential computation is already a mature research subject. A witness is the rich algebraic theories based on regular expressions and the associated regular algebra \cite{kle56,sal66,conway71,kuich/salomma85,ste87b,koz91}. Recent extensions of regular algebra to network algebra \cite{ste86,stefNA} show a broader area of applications and deep connections with classical mathematics, especially via the particular instance provided by trace monoidal categories \cite{jsv96,sel11}. The present approach is an attempt to extend these formalisms to open, distributed, interactive programs.

There are a few models for parallel/distributed computation based on regular expressions. We mention two of them: regular expressions for Petri nets \cite{ga-ra92} and for timed automata \cite{acm02}. Both are based on renaming and intersection, two questionable operations.

Our work on this subject started with the exploration of \textit{space-time duality} and its role in organizing the space of interactive computation:
(1) A space-time invariant model extending \textit{flowcharts} was shown in \cite{ste04,ste06a}. (2) A space-time invariant extension of (structured) \textit{while programs} was presented in \cite{dr-st08a}. (3) An enriched version supporting \textit{recursion} was presented in \cite{pss07}. (4) Space-time invariant (2-dimensional) \textit{regular expressions} are presented in \cite{bps13,ba-st14b,ba-st14a} and in the current paper. (5) Verification methods lifting \textit{Floyd and Hoare logics} to 2-dimensions have been described, too. (6) A notion of \textit{refinement}, based on this space-time invariant model, was introduced in \cite{dlpss12,dpss14}.

To conclude: the proposed research programme on clarifying the structure of self-assembling (2-dimensional) tiles is not only of general interests, but it is also very important in understanding open, distributed, interactive computation.